\documentclass[11pt,a4paper]{article}
\usepackage[all]{xy}
\usepackage{amsmath,amsfonts,amssymb,amsthm,amscd,pb-diagram,pb-xy}
\usepackage{graphicx}
\usepackage{epsfig}
\usepackage{caption}

\usepackage{color}  

\definecolor{red-}{rgb}{0.8,0.0,0.0}

\definecolor{green-}{rgb}{0.0,0.7,0.0}

\definecolor{blu-}{rgb}{0.0,0.0,1.0}

\newcommand{\Reali}{\ensuremath{\mathbb{R}}}
\newcommand{\Euclideo}{\ensuremath{\mathbb{E}}}
\def\vth{\vartheta}
\def\vph{\varphi}
\def\A{\mathcal A\/}
\def\B{\mathcal B\/}

\def\H{\mathcal H\/}
\def\K{\mathcal K\/}
\def\M{\mathcal M\/}

\def\Q{\mathcal Q\/}
\def\P{\mathcal P\/}

\def\S{\mathcal S\/}

\def\J{J_1\/(\M)}

\def\JS{J_1\/(\S)}

\def\MA{V(\A)}
\def\MB{V(\B)}

\def\V{\mathcal V\/}
\def\KinH{^{\bf h} \K}
\def\velort{{\bf V}^{\perp}\/}
\def\velortS{{\bf V}^{\perp}_{\S}\/}
\def\velortB{{\bf V}^{\perp}_{\B}\/}

\def\={\, = \,}
\def\de#1/de#2{\frac{\partial {#1}}{\partial {#2}}}
\def\De#1/de#2{\dfrac{\partial {#1}}{\partial {#2}}}

\def\q#1{q\,^{#1}}

\def\dq#1{\dot{q}\,^{#1}}

\def\x#1{x\,^{#1}}

\def\dx#1{\dot{x}\,^{#1}}

\def\ES{\varepsilon_{\S}}
\def\EB{\varepsilon_{\B}}
\def\pl{{\bf p}_L}
\def\pr{{\bf p}_R}

\begin{document}

\title{Geometric characterization of frictional impacts by means of breakable kinetic constraints}

\author{Stefano Pasquero\thanks{The author thanks the support by University of Parma and by the Italian National Group of Mathematical Physics
		(GNFM-INdAM)}\\
        Departments of Mathematics, Physics and Computer Sciences \\
        University of Parma \\
        Parco Area delle Scienze 53/a, 43124 PARMA -- Italy \\
        E-mail: stefano.pasquero@unipr.it
        }
\date{}
\maketitle
\begin{abstract}
\noindent In the context of geometric Impulsive Mechanics of systems with a finite number of degrees of freedom, we model the roughness of a unilateral constraint $\S$ by introducing a suitable instantaneous kinetic constraint $\B\subset \S$. A constitutive characterization of $\B$ based only on the geometric properties of the setup and on the dry friction laws can then be introduced to model the frictional behavior of $\S$ in an impact of the system. Such a model restores determinism and avoids the analysis of frictional forces in the contact point, with all its associated theoretical problems of causality. Three examples of increasing complexity, showing a natural stick--slip behavior of the impact, are presented.
\end{abstract}

\section*{Introduction}

Classical ``smooth'' Mechanics and classical impulsive Mechanics both take their origin and foundation in Newton's Second Law (NSL) ${\bf F} \= m {\bf a}$ applied to a material point $P$, so unavoidably the core ideas on which to base their studies are very similar if not the same. However, not all the common logical bases of these ideas have the same applicability and are equally fruitful when applied in the different contexts of smooth or impulsive Mechanics.

As an example, let us consider the causal structure of the NSL: it is well known that, in the equation ${\bf F} \= m {\bf a}$, the force ${\bf F}$ plays the role of the cause of the acceleration ${\bf a}$, that is the effect. This implies that, for the Second Law to be fruitful, the force should be a known data of the problem (see, e.g. \cite{Peres}). Then, in the context of strictly mechanical phenomena, ${\bf F}$ is a known function ${\bf F} \= {\bf F}(t, P, {\bf v})$ of time, position and velocity of the point, that allows to determine the (acceleration and then possibly the) motion of $P$. This line of thought is pursued by the introduction of the various kinds of active forces (elastic, gravitation, Coulomb, Lorentz and so on...).

However, a similar procedure cannot be applied straightforwardly in the context of impulsive Mechanics. The impulsive version of the NSL  ${\bf P} \= m \Delta {\bf v} \= m ({\bf v}_R - {\bf v}_L)$ can be rewritten with the correct ``causal'' structure in the form ${\bf P}(t, P, {\bf v}_L)  \= m {\bf v}_R$, but, for several reasonable causes (see e.g. \cite{pasquero2018hard}), a satisfactory analysis of the possible forms of the active impulse  ${\bf P}$ is not available in literature. Then, concerning the causal aspect, smooth and impulsive Mechanics start from the same point but diverge rapidly.

\smallskip

A similarity in smooth and impulsive Mechanics occurs regarding the approach to reactive forces and reactive impulses in presence of constraints acting on the system. In fact reactive forces and impulses are by nature unknowns of the mechanical problem, and, in order to restore determinism, that is one of the cornerstone of Classical Mechanics, it is necessary to introduce a constitutive characterization of the constraints, that is an {\it a priori} prescription of what (kind of) forces or impulses the constraint can or cannot exerts. The methodology at the basis of the choice of this prescription is common for smooth and impulsive Mechanics, and it is common also for simple mechanical model (e.g. material points) and for more refined ones (e.g. holonomic systems): it can be based on the geometric properties of the system (e.g. orthogonal or parallel decomposition of kinematic or dynamical quantities with respect to the constraint), and on conservation or dissipative laws (e.g. conservation of kinetic energy). This is the reason why a preliminary analysis of the geometry of constrained systems is essential, both for smooth and impulsive Mechanics, to introduce the several possible choices of constitutive characterization.

However, once again the results of the same line of thought applied to smooth or impulsive systems can be very different, and this can happen both for simple systems (e.g. material points) and holonomic systems (e.g. mechanical systems with a finite numer of degrees of freedom): for instance, the Amontons--Coulomb--Morin laws give a well defined constitutive characterization for dry friction acting on a single constrained material point, but they cannot be straightforwardly applied to holonomic systems, as shown by the Painlev\'e paradox. Moreover, the modeling of friction in impulsive Mechanics of holonomic systems is, as yet, an extremely challenging subject, about which several very different models can be found in literature (a very concise survey about this is presented in \cite{GILARDI2002}).

\smallskip

In this paper we present a model for the frictional behavior of unilateral constraints in impulsive Mechanics of holonomic systems. The model is framed in the context of the so--called {\it event driven} approach of impacts having a single contact point of the system with the constraint. The main result consists in showing that the frictional reaction in the impact of a holonomic system  with a unilateral constraint can be modeled with the constitutive characterization of a suitable breakable instantaneous kinetic constraint.

The rationale of this model is based on two parts: the first is the geometric structure of the configuration space--time bundle of the system, together with their constraint subbundles; the second is the Amontons--Coulomb--Morin empirical laws for dry friction applied to a constrained material point $P$
\begin{subequations}\label{AC}
	\begin{eqnarray}\label{ACS}
		\hskip -1.3truecm \| {\bf \Phi}^s_T \| \, &\le \, \mu_s \|
		{\bf \Phi}^s_N \|
	\end{eqnarray}   \vskip-0.7truecm
	\begin{eqnarray}\label{ACD}
		{\bf \Phi}^d_T \, \, = \, -  \mu_d  \, \|{\bf \Phi}^d_N\| \,
		\dfrac{{\bf v}}{\| {\bf v} \|} \, ,
	\end{eqnarray}
\end{subequations}
where ${\bf \Phi}^s_T$ and ${\bf \Phi}^d_T$ are the static and
dynamic dry friction forces respectively, $\mu_s \ge \mu_d$ are
the static and dynamic friction coefficients respectively (possibly depending on time and position),
${\bf \Phi}^s_N$ and ${\bf \Phi}^d_N$ are the static and dynamic loads
respectively, both determinable using the Newton's Laws, and
$\dfrac{{\bf v}}{\| {\bf v} \|}$ is the unitary vector of the velocity.

The model has two important characteristics: being based on a constitutive characterization of a constraint, it structurally restores determinism; moreover, it does not require any kind of study of active or reactive forces arising in the contact point, so that it does not involve any causal aspect that could arise with the use of NSL in an impulsive context.

\smallskip

After a preliminary definition of regular geometric impulsive mechanical system (RGIMS), in order to model the impact of the system with a rough unilateral constraint $\S$ of codimension $1$ we introduce the instantaneous kinetic constraint $\B$ expressing the vanishing of the velocity of the point of the mechanical system in contact with $\S$. Then the geometric properties of the RGIMS determine, together with a natural direction orthogonal to $\S$, also a direction tangent to $\S$ and orthogonal to $\B$. Every left--velocity $\pl$ of the system determining an impact with $\S$ can then be naturally split at first into   a part tangent to $\S$ and a part orthogonal to $\S$ and, secondly, the tangent part can be in its turn split in a part tangent to $\B$ and a part orthogonal to $\B$. These three objects are the ingredients that allow to assign a constitutive characterization that express both the inelasticity and the roughness of $\S$. As the concept of inelasticity of $\S$ is quite well known, taking into account the classification of constraints presented in \cite{Pasquero2018RMUP} the roughness of $\S$ can be modeled by requiring the breakability of $\B$, that is, by requiring that the constitutive characterization of the reaction exerted by $\B$ is split into two or more different possibilities depending on the geometric and/or kinetic data of the impact. The breaking rule of the characterization of $\B$ is inspired by the Morin--Amontons--Coulomb laws of friction.

\smallskip

In Section 1 we define the regular geometric impulsive mechanical systems (RGIMS), and briefly recall their geometric structure, particularly fit to model Impulsive Mechanics. The section contains well known material, as presented for example in \cite{Pasquero2006}, and it appears only to make the paper self consistent.

In Section 2 we define the general concept of constitutive characterization of the impulsive constraint. We distinguish between ideal ones, non ideal ones without friction, and non ideal ones with friction. The first two types are once again known in literature, the last one is introduced with the new ideas heavily based on both the geometric stucture of RGIMS and on the Morin--Amontons--Coulomb laws for dry friction. The nature itself of the characterization shows the possible stick--slip behavior of the impacting system.

In Section 3 we present three examples of RGIMS having a frictional impact with increasing computational hardship: a material point, a rigid disk, a rigid rod, all impacting with a rough horizontal floor. The examples show the reasoning and clearness of results but also the increasing complication in their interpretation.

The list of references is based on the minimality criterion of making the paper
self-consistent and then is focussed on the specific approach given by the geometric
framework of RGIMS. A different choice, even if restricted to works pertaining only frictional impacts, could draw away the attention of the reader from the peculiar approach of
this paper. However, the reader interested in classical approach to contact/impact mechanics
can find foundations, main results and  huge bibliographies in \cite{Johnson, Stronge, Brogliato}. Different approaches to frictional impacts can be found in  \cite{pfeiffer1995impacts,Torre1994,Steward2000}.

\section{Preliminaries}

In this section we recall, also in order to fix notation, the structures and the principal
properties of the geometric environment of Impulsive Mechanics. The section contains standard
material that is presented, as briefly as possible, to keep the paper self--contained. For this reason, we do not delve into the technicalities of the unilateral aspects of the constrained system. More ample
discussions can be found in literature (see for example
\cite{Saunders,MassPaga1997,Pasquero2004,Pasquero2005uni,Pasquero2006,Pasquero2008,pasquero2018MIl}).

\subsection{Geometric structure of RGIMS}

A Regular Geometric Impulsive Mechanical System (RGIMS) is a 5-uple $(\M, \Phi, \S, \A, \B)$
where:
\begin{itemize}
	\item[1)] $\M$ is the {\it configuration space--time} bundle of the RGIMS. It is a
	$(n+1)$--dimensional fiber bundle $\pi_t: \M \to \Euclideo$, where $\Euclideo$ is the $1$-dimensional euclidean space. All the fibers of $\pi_t: \M \to \Euclideo$ are
	diffeomorphic to an $n$--dimensional manifold $\Q$, usually thought of as the {\it configuration
		space\ } of the system. We will refer $\M$ to  {\it admissible} coordinate systems $(t, x^1,
	\ldots, x^n)$ having the  {\it absolute time function} $t:\Euclideo \to \Reali$, a global cartesian coordinate on $\Euclideo$, as first coordinate. Once the coordinate $t$ is fixed, we can equivalently consider as space--time bundle  the fibered manifold $t: \M \to \Reali$;

	\item[2)] $\Phi$  is the {\it vertical metric} of $\M$. It is a differentiable positive definite
	scalar product $\Phi : V(\M) \times_{\M} V(\M) \to \Reali$, where $V(\M)$ is the vertical vector
	bundle of the configuration space--time and $\times_{\M}$ denotes the usual fiber product of
	bundles on $\M$. As usual, we denote by $g_{ij}$ the so-called {\it mass matrix} of the system, that is the set of functions
	\begin{eqnarray*}
		g_{ij}(p) \, = \, \Phi \, \left( \left(\De{}/de{x^i}\right)_p , \left(\De{}/de{x^j}\right)_p
		\right), \qquad p \in \M \, ,
	\end{eqnarray*}
	and we recall that the $g_{ij}$ take intrinsically into account the mass properties of the system
	(see \cite{MassPaga1997});

	\item[3)] $\S$ is the {\it contact constraint bundle} of the RGIMS. It is a subbundle $t:\S\,\to
	\Reali$ of the the bundle $\M$, with $\S$ of codimension $1$. As with the bundle $t:\M\,\to
	\Reali$, the bundle $t:\S\,\to \Reali$ admits its own admissible coordinates $(t,\q{1},\ldots,
	\q{n-1})$;

	\item[4)] $\A$ is the {\it permanent kinetic constraint bundle} of the RGIMS. It is an affine
	subbundle $a: {\A} \,\to \J$ of the affine bundle $\pi : \J \to \M$, where $\J$ is the first jet
	extension of the configuration space--time. The manifold $\A$ is globally fibered over $\M$ (and
	then over the real line), and $\A$ has dimension $(n+ s+1), s \le n$. The bundle $\A$ will be referred
	to admissible coordinates $(t,\x{i},\dot\xi^{A})$. However, in the following, with the exception of the example in \ref{ExampleRGIMS}, $\A$ will always coincide with the whole $\J$;

	\item[5)] $\B$ is the {\it instantaneous kinetic constraint bundle} of the RGIMS. It is an affine
	subbundle $b: {\B} \,\to \JS$ of the affine bundle $\pi : \JS \to \S$, where $\JS$ is the first jet
	extension of the contact bundle $\S$. The manifold $\B$ is globally fibered over $\S$, and has
	dimension $(n+m), m \le n$. The bundle $\B$ will be referred to admissible coordinates
	$(t,\q{\eta},\dot\theta^{\Upsilon})$.

\end{itemize}

\medskip

The knowledge of a RGIMS determines several geometric objects that are relevant in the description
of the mechanic behavior of the system in study. The following spaces and properties, some of them already mentioned in the definition of RGIMS and naturally introduced by the
assignment of a RGIMS, complete the geometric setup:

\begin{itemize}
	\item[1)] the set
	$V(\M) \= \left\{ {\bf X} \in T(\M)  \, \, | \,\,  <{\bf X}, d t> \= 0 \right\}$
	is a vector subbundle of $T(\M)$ called the {\it vertical space of} $t:\M\,\to
	\Reali$. Its elements are called {\it space--like} vectors and, in admissible coordinates, they have the local expression ${\bf X} \= X^i \left(\De{}/de{x^i}\right)$.

	The set
	$\J \= \left\{ {\bf p} \in T(\M)  \, \, | \,\, <{\bf p}, d t> \= 1
	\right\}$
	is the affine subbundle of $T(\M)$ called the first jet extension of $\M$. Its elements  are called {\it time--like} vectors and, in admissible coordinates, they have the local expression ${\bf p} \= \left(\De{}/de{t}\right)_p  +  X^i \left(\De{}/de{x^i}\right)_p$. The bundle $\J$ has a natural structure of affine bundle modeled on the vector bundle $V(\M)$.

	Analogous sets $V(\S), \JS$ can be defined for the contact constraint bundle $t:\S\,\to
	\Reali$;

	\item[2)] the set
	$\MA \= \left\{ {\bf A} \= {\bf a}_1 \, - \, {\bf a}_2 \, \, | \,\,  {\bf a}_1 , {\bf a}_2 \in \A
	\right\}$
	is the {\it vertical space of $\A$}. Due to the affine structure of $\A$, it is a vector subbundle
	of $V(\M)$ fibered over $\M$;

	\item[3)] the set
	$
	\MB \= \left\{ {\bf B} \= {\bf b}_1 \, - \, {\bf b}_2 \, \, | \,\,  {\bf b}_1 , {\bf b}_2 \in \B
	\right\}$
	is the {\it vertical space of $\B$}. Due to the affine structure of $\B$, it is a vector subbundle
	of $V(\S)$ fibered over $\S$;

	\item[4)] being $V(\S), \MA$ subbundles of $V(\M)$, they inherit a scalar product induced by the vertical metric $\Phi$ of $V(\M)$. The same then holds for $\MB \subset V(\S)$;

	\item[5)] due to the affine structure of $\J$, every global section ${\bf h} : \M \to \J$ determines
	a ``vectorialization'' of $\J$, that is a fibered bijection $\Delta_{\bf h}:\J \to V(\M)$ such
	that $\Delta_{\bf h}({\bf p}) \= {\bf p} - {\bf h}$;

	\item[6)] the set $\H_{\S}$ is the set of the global sections ${\bf h} : \M \to \J$ such that ${\bf h}_{\rfloor_p} \in \JS \,\, \forall \, p\in \S$. They are represented by
	vector fields having the local form ${\bf h} \, = \, \De{}/de{t} + H^i(t, \x{j}) \De{}/de{\x{i}}$ and whose restriction to $\S$ is tangent to $\S$. Then, in particular, their one--parameter group of transformations maps the constraint $\S$
	into itself;

	\item[7)] the set $\H_{\A}$ is the subset of $\H_{\S}$ formed by the sections (tangent to $\S$ and)
	respecting the conditions determined by $\A$;

	\item[8)] the set $\H_{\B}$ is the subset of $\H_{\S}$ formed by the section (tangent to $\S$ and)
	respecting the conditions determined by $\B$;

	\item[9)] the set $\H_{\A\cap\B}$ is the subset of $\H_{\S}$ formed by the section (tangent to $\S$
	and) respecting the whole set of conditions determined by $\A$ and $\B$.
\end{itemize}

The geometric setup of a RGIMS is synthesized by the following diagram:
\begin{equation}\label{diagramma}
	\xymatrix @R=15pt @C=10pt{%
		&&V(\B) \ar |(0.4){}[rr] &&V(\S) \ar |(0.5){i_*}[rrr] \ar |(0.7){\pi} [ddddl] &&&V(\M) \ar
		|(0.7){\pi} [ddddl] &&
		V(\A) \ar |(0.4){}[ll] &&\\ \\ %
		\B \ar |(0.4){}[rr] &&\JS \ar |(0.7){i_*}[rrr] \ar |(0.5){\pi} [ddr] &&&\J \ar |(0.5){\pi} [ddr]
		\ar@/^1pc/@{<.>} |(.5){\Delta_{\bf h}}[uurr] &&&
		\A \ar |(0.4){}[lll] && \\ \\ %
		&&&\S \ar |(0.5){i}[rrr] \ar |(0.5){t} [dd] &&&\M \ar@/^.8pc/@{..>} |(.4){\bf h}[uul] \ar |(0.5){t} [dd] &&&&&&&\\ \\ %
		&&&\Reali \ar@{=}[rrr] &&&\Reali  &&&&&&&\\ \\ %
	}%
\end{equation}

Of course, both $\A$ and $\B$ could be trivial, so that $\A\equiv\J$  and/or $\B\equiv\JS$. Note
moreover that the three constraint bundles can be assigned in their cartesian form
\begin{eqnarray*}
	\begin{array}{ll}
		\S: s(t, x^i) \= 0; \\ %
		\A: a_{\beta}(t, x^i, \dx{i}) \= 0 \qquad &\beta=1,\ldots, (n-s); \\ %
		\B: b_{\gamma}(t, q^{\alpha}, \dq{\alpha}) \= 0 \qquad &\gamma=1,\ldots, (n-1-m).
	\end{array}
\end{eqnarray*}

To conclude the construction, let us recall that we focus our attention on the impact of the
system with the constraint $\S$, so that we are interested only in those configurations of the
system belonging to $\S$. Therefore, a more proper setup to pursue a geometric analysis of the
impact consists in the pair of pull--back bundles $\pi: i^*(\J) \to \S$ and $\pi: i^*(V(\M)) \to
\S$ over $\S$  (with their subbundles $\B\subset\JS\subset i^*(\J), \, i^*(\A) \subset i^*(\J)$
and $V(\B)\subset V(\S) \subset i^*(V(\M)), \, i^*(V(\A))\subset i^*(V(\M))$).

Later on, we will pass over this last technicality and we will continue to refer to diagram
(\ref{diagramma}). Moreover, with a mild abuse of language, we will not distinguish the subbundles
and their images through immersion. Then, for brevity, we will consider $\B\subset\JS\subset\J, \,
\A\subset \J$ and similarly, $V(\B)\subset V(\S) \subset V(\M), \, V(\A)\subset V(\M)$.

\subsection{Mechanical aspects of RGIMS}

The roles played by the bundles depicted in diagram (\ref{diagramma}) in the context of
(Impulsive) Mechanics is well known (\cite{MassPaga1997,Pasquero2005uni}) and are summarized as
follow:

\begin{itemize}
	\item[1)] every global section ${\bf h} : \M \to \J$, whose integral lines determine a 1 to 1 correspondence between the points of every chosen pair of fibers of $\M$, can be considered a frame of reference for the system. In particular, the set $\H_{\S}$ represents the set of the possible {\it frames of rest of $\S$}. The sets $\H_{\A}, \H_{\B}$ and $\H_{\A\cap\B}$ represent the restrictions of the set $\H_{\S}$ to those frames
	respecting the kinetic conditions determined by $\A, \B$ and $\A\cap\B$ respectively;

	\item[2)] the time--like vectors ${\bf p} \in \J$ of the first jet extension of the configuration space--time are the absolute velocities of the mechanical system.

	The space--like vectors ${\bf X} \in V(\M)$ have a dual nature:  once a frame of reference ${\bf h}$ is assigned, the vertical vector ${\Delta_{\bf h}}({\bf p}) \= {\bf p} - {\bf h}_{\rfloor_{\pi({\bf p})}} \in V(\M)$ is the relative (to ${\bf h}$) velocity
	corresponding to the absolute velocity ${\bf p}$.
	In particular, the function
	\begin{eqnarray}\label{EnergiaCinetica}
		\KinH : \J \to \Reali \qquad {\textrm s.t.} \qquad \KinH({\bf p}) \= \dfrac{1}{2} \Phi({\bf p - h, p - h})
	\end{eqnarray}
	is the kinetic energy of the system with respect to the frame ${\bf h}$.

	However, since  $V(\M)$ is the modelling vector bundle
	of the affine bundle $\J$, the difference ${\bf X} \= {\bf p}_2 - {\bf p}_1$ of two absolute velocities is an element of $V(\M)$. Then $V(\M)$ can also be viewed as the space of all possible jumps of absolute velocities of the system. This is the mathematical basis of the ``impulsive'' relation
	\begin{eqnarray}\label{relazione impulso}
		{\bf p}_R \= {\bf p}_L + {\bf I},
	\end{eqnarray}
	where ${\bf p}_L,{\bf p}_R \in \J$ are the absolute velocities of the system before and after the
	impact respectively and ${\bf I} \in V(\M)$ is an impulse, possibly the reactive impulse due to the constraints acting on the system;

	\item[3)]  the subbundle $\A \subset \J$ represents the admissible velocities of the system when
	kinetic restrictions are required on the velocities of the RGIMS before and after the impact
	(justifying then the expression {\it permanent} kinetic constraint bundle). Its corresponding
	vertical bundle $V(\A)$ is the space of the possible impulses acting on the system and respecting
	the kinetic restrictions given by $\A$;

	\item[4)]  the subbundle $\B \subset \JS$ represents the possible kinetic restrictions required on
	the velocities of the RGIMS at the moment of the impact (justifying then the expression {\it
		instantaneous} kinetic constraint bundle);
\end{itemize}

\noindent \underline{REMARK}. The term ``regular'' in the definition of RGIMS would require some explanation, for two main reasons: the first is the possibility of defining non regular GIMS, and the corresponding physical meaning (especially referred to multiple impacts); the second is because it is necessary to adjoin the ulterior condition $\A \cap \B \ne \emptyset$ to ensure that the RGIMS has a clear physical meaning. Since these aspects will be not relevant for the following, the reader is referred to \cite{Pasquero2018RMUP,Pasquero2005gauss,Pasquero2006} for details.

\subsection{Example of RGIMS}\label{ExampleRGIMS}

A ball of mass $m$ and radius $R$ moves in contact with a horizontal plane and impacts with a vertical wall.
The bundle $\M$ is a $6$--dimensional manifold that can be  locally described by coordinates
$(t,x,y,\psi,\vth,\vph)$, with $x,y$ coordinates of the center of the ball, and $(\psi,\vth,\vph)$
the Euler angles relative to the ball. The vertical metric is given by the matrix functions
\begin{eqnarray*}
	g_{ij} \= \left(
	\begin{array}{ccccc}

		m & 0 & 0 & 0 & 0 \\
		0 & m & 0 & 0 & 0 \\
		0 & 0 & A & 0 & A\cos\vartheta \\
		0 & 0 & 0 & A & 0 \\
		0 & 0 & A\cos\vartheta & 0 & A
	\end{array}
	\right)
\end{eqnarray*}
where $A$ is the momentum of inertia of the ball. The contact constraint bundle $\S$ is described
by the equation $y\=R$, so that $\S$ is locally described by coordinates $(t,x,\psi,\vth,\vph)$. A
permanent kinetic constraint bundle $\A$ can be introduced, for instance, by requiring the pure rolling of the
ball on the horizontal plane, so that $\A$ is described by the equations:
\begin{eqnarray*}
	\left\{
	\begin{array}{l}
		\dot{x} - R \dot{\vth}\sin\psi + R \dot{\vph} \sin\vth\cos\psi \= 0 \\
		\dot{y} + R \dot{\vth}\cos\psi + R \dot{\vph} \sin\vth\sin\psi \= 0 .
	\end{array}
	\right.
\end{eqnarray*}
An instantaneous kinetic constraint bundle $\B$ can be introduced by requiring the pure rolling of
the ball on the vertical wall, so that $\B$ is described by the equations:
\begin{eqnarray*}
	\left\{
	\begin{array}{l}
		\dx{} + R \dot{\psi} + R \dot{\vph} \cos\vth \= 0 \\
		\dot\vth \cos\psi  + \dot\vph \sin\vth\sin\psi  \= 0
	\end{array}
	\right.
\end{eqnarray*}

\section{Constitutive characterizations for RGIMS}

In this section we introduce, in the general context of RGIMS, the decomposition of absolute velocities in parts tangent and orthogonal to the constraints. In particular, the orthogonal components constitute the starting ingredients to discuss how to assign an (ideal or non ideal)
constitutive characterization for the constraints in the RGIMS.

\subsection{Orthogonal velocities and constitutive characterization}

The simultaneous presence of the vertical metric $\Phi$ and the inclusions $V(\B)\subset V(\S)
\subset V(\M), V(\A)\subset V(\M)$ determine the splits
\begin{eqnarray}\label{split vettoriale}
	\begin{array}{lcl}
		V(\M) &\=& V(\S) \oplus_{\M} V(\S)^{\perp} \\
		V(\M) &\=& V(\A) \oplus_{\M} V(\A)^{\perp} \\
		V(\S) &\=& V(\B) \oplus_{\S} V(\B)^{\perp}
	\end{array}
\end{eqnarray}
and the corresponding six possible projection operators (that with obvious notation will be denoted with the
letter $\V$ with suitable specifications, such as for examples, $\V^{\perp}_{\A}$ or
$\V^{\|}_{\B}$). In their turn, the splits (\ref{split vettoriale}) determine the splits
\begin{eqnarray}\label{split affine}
	\begin{array}{lcl}
		\J &\=& \JS \oplus_{\M} V(\S)^{\perp} \\
		\J &\=& \A \oplus_{\M} V(\A)^{\perp} \\
		\JS &\=& \B \oplus_{\S} V(\B)^{\perp}
	\end{array}
\end{eqnarray}
(with a mild abuse of notation relative to the sign $\oplus$) and the corresponding six possible
projection operators (that will be denoted with the letter $\P$ with suitable
specifications, such as for examples, $\P^{\perp}_{\A}$ or $\P^{\|}_{\B}$). In particular, note that, in presence of a proper instantaneous kinetic constraint $\B \subset \JS$, we have the split
\begin{eqnarray}\label{TripliceSpezzamento}
	\J \= \B \oplus_{\S} V(\B)^{\perp} \oplus_{\M} V(\S)^{\perp} \, .
\end{eqnarray}

Needless to say, the splits (\ref{split vettoriale},\ref{split affine}) have an absolute (in the
sense of ``independent of any frame of reference'') meaning. However, it is a well known result
(\cite{Pasquero2005uni,Pasquero2006}) that
\begin{eqnarray}\label{ProiettoriOrtogonali}
	\begin{array}{lcll}
		\P^{\perp}_{\S}({\bf p}) &\=& \V^{\perp}_{\S}({\bf p} - {\bf h}) \quad &\forall \, {\bf p} \in \J
		, \forall \, {\bf h} \in \H_{\S} \\ \\ %
		\P^{\perp}_{\A}({\bf p}) &\=& \V^{\perp}_{\A}({\bf p} - {\bf h}) \quad &\forall \, {\bf p} \in \J
		, \forall \, {\bf h} \in \H_{\A} \\ \\ %
		\P^{\perp}_{\B}({\bf p}) &\=& \V^{\perp}_{\B}({\bf p} - {\bf h}) \quad &\forall \, {\bf p} \in \JS
		, \forall \, {\bf h} \in \H_{\B}.
	\end{array}
\end{eqnarray}
Note that these relations are in general not verified without the restriction posed on the frames of reference. Moreover, no analogous relations hold for the parallel projectors.

Relations (\ref{ProiettoriOrtogonali}) give then an absolute meaning to the action of the orthogonal projectors $\P^{\perp}$, whose results on an absolute velocity ${\bf p}$
is a vertical vector that can be thought of as the {\it absolute} orthogonal component of
${\bf p}$ with respect to the corresponding constraint. Absolute orthogonal velocities will be
denoted with ${\bf V}^{\perp}$ (and specifying the corresponding constraint).

\bigskip

A {\it constitutive characterization} for a RGIMS is a rule ${\bf I}:\A \to V(\M)$ assigning to each admissible
``left velocity'' ${\bf p}_L$ of the system a corresponding ``reactive impulse'' ${\bf I}(\pl)$. The mechanical behavior of the system (and the restore of the determinism) is
therefore obtained by the relation (\ref{relazione impulso}), expressing the ``right velocity'' $\pr= \pl + {\bf I}(\pl)$ as a function of the ``left velocity''.

Due to the wide generality of the definition, essentially based on the requirement of restoring
the mechanical determinism, the choice of a
constitutive characterization can be motivated by the most various assumptions. However, among the
weak requirement of the mathematical admissibility, constitutive characterizations should also
meet physically meaningful criterions, of both theoretical and experimental type.

\subsection{Non--frictional constitutive characterizations}

Several constitutive characterizations based on the splits presented above have already been presented for different kinds of RGIMS.

\begin{itemize}
	\item if $\A = \J, \B = \S$, then the only orthogonal component of $\pl$ is ${\bf V}^{\perp}_{\S}(\pl)$. The choice (see \cite{Pasquero2005uni})
	\begin{eqnarray}\label{caratt ideale 1}
		{\bf I} : \J \to V(\M) \quad {\rm s.t.} \quad {\bf I}(\pl) \= - \, 2 \, {\bf V}^{\perp}_{\S}(\pl)
	\end{eqnarray}
	determines a right velocity
	\begin{eqnarray*}
		\pr \= \pl - \, 2 \, {\bf V}^{\perp}_{\S}(\pl) 	\= \P^{\|}_{\S}(\pl ) - \, {\bf V}^{\perp}_{\S}(\pl).
	\end{eqnarray*}
	It corresponds to a complete reflection of the orthogonal component ${\bf V}^{\perp}_{\S}(\pl)$. Since the reactive impulse does not have parts tangent to $\S$ and this choice satisfies also the requirement of invariance of the kinetic energy of the system
	with respect to all the frames in the class $\H_{\S}$ (that is the widest class of frames of reference for which we can ask for the conservation of kinetic energy in this case), then (\ref{caratt ideale 1}) is a non--frictional ideal constitituve characterization for $\S$;

	\item if $\A = \J, \B = \S$ we can choose (see \cite{Pasquero2008})
	\begin{eqnarray}\label{caratt non_ideale 1}
		{\bf I} : \J \to V(\M) \quad {\rm s.t.} \quad {\bf I}(\pl) \= - \, (1+\ES) \, {\bf V}^{\perp}_{\S}(\pl)
	\end{eqnarray}
	with $\ES \in [0,1)$ is a restitution coefficient expressing ``how much'' of the velocity orthogonal to the contact constraint is restituted to the system. Then we obtain
	\begin{eqnarray*}
		\pr \= \P^{\|}_{\S}(\pl ) - \, \ES \, {\bf V}^{\perp}_{\S}(\pl).
	\end{eqnarray*}
	The coefficient $\ES$ is clearly related to the more or less elastic response of the contact constraint and, with due caution to the dependence on the frame of reference, to the dissipation of the kinetic energy of the system. Since the reactive impulse does not have parts tangent to $\S$, this is a non--frictional non--ideal constitutive characterization for $\S$.
\end{itemize}

\subsection{Model for frictional constitutive characterization}

The presence of friction in $\S$ presupposes a tangent (to $\S$) component of the impulsive reaction ${\bf I}(\pl)$. Taking into account the split (\ref{TripliceSpezzamento}), we can introduce  a proper instantaneous kinetic constraint  $\B \subset \JS$ expressing the vanishing of the velocity of the point of the system in contact with the unilateral constraint $\S$. Then every absolute velocity $\pl$ can be split in the form
\begin{eqnarray*}
	\pl \=  \P^{\|}_{\B}(\pl) + {\bf V}_{\B}^{\perp}(\P^{\|}_{\S}(\pl)) + {\bf V}_{\S}^{\perp}(\pl) \, .
\end{eqnarray*}
Being the orthogonal components ${\bf V}_{\B}^{\perp}(\P^{\|}_{\S}(\pl))$ tangent to the contact constraint $\S$, every constitutive characterization introducing a component of the reactive impulse parallel to this velocity can be thought of as a frictional characterization. Quite naturally, parallelizing (\ref{caratt non_ideale 1}), we can introduce of a second coefficients $\EB \in [0,1)$ and a very simple non ideal  characterization expressing a ``double'' loss of returned velocity of the system in the orthogonal to $\S$ and in the orthogonal to $\B$ directions. We can then set
\begin{eqnarray*}
	{\bf I}(\pl) \= - \, (1+\ES) \, {\bf V}^{\perp}_{\S}(\pl) - \, (1+\EB) \, {\bf
		V}^{\perp}_{\B}(\P^{\|}_{\S}(\pl))
\end{eqnarray*}
that determines the right velocity
\begin{eqnarray*}
	\begin{array}{lcl}
		\pr &\=& \P^{\|}_{\B}(\pl ) - \, \ES \, {\bf V}^{\perp}_{\S}(\pl) - \, \EB \, {\bf
			V}^{\perp}_{\B}(\P^{\|}_{\S}(\pl)).
	\end{array}
\end{eqnarray*}

However, although this characterization presents a frictional part, it is more similar to a dissipative characterization rather then a frictional one. If we want to relate the tangent impulsive reaction to a more common idea  of friction on the contact constraint, $\EB$ must play a more specific role in a ``Coulomb''--like constitutive characterization. Adopting the shorter notation ${\bf
	V}^{\perp}_{\B}(\P^{\|}_{\S}(\pl)) = {\bf
	V}^{\perp}_{\B}(\pl))$ and omitting the argument, we set
\begin{eqnarray}\label{legge d'attrito}
	\EB \= - 1 \, + \, \dfrac{\min\left\{\|\velortB \|, \mu_s \|\velortS\| \right\}}{\|\velortB\|} ,
\end{eqnarray}
where  $\mu_s$ is the static friction coefficient of the Morin--Amontons--Coulomb dry friction law. This implies that
\begin{eqnarray}\label{impulso_attrito}
	{\bf I}(\pl) \= \left\{
	\begin{array}{lclcl}
		- \, (1+\ES) \, {\bf V}^{\perp}_{\S} \, -  \, {\bf
			V}^{\perp}_{\B} &\,& \,& {\rm if} \, \|{\bf
			V}^{\perp}_{\B}\| \le  \mu_s \| {\bf V}^{\perp}_{\S} \| \\ \\
		- \, (1+\ES) \, {\bf V}^{\perp}_{\S} \, - \, \dfrac{\mu_s \|\velortS\|}{\|\velortB\|} \, {\bf
			V}^{\perp}_{\B} &\, & \,& {\rm if} \,  \|{\bf
			V}^{\perp}_{\B}\| >  \mu_s \|{\bf V}^{\perp}_{\S}\|
	\end{array}
	\right.
\end{eqnarray}
so that (see \cite{Pasquero2018RMUP}) $\B$ becomes a breakable instantaneous kinetic constraint for the system. The right velocity corresponding to (\ref{impulso_attrito}) is then
\begin{eqnarray}\label{velocita_finale_attrito}
	\hskip-0.4truecm
	\pr \= \left\{
	\begin{array}{ll}
		\P^{\|}_{\B}(\pl ) - \, \ES \, {\bf V}^{\perp}_{\S}(\pl) & \,\, {\rm if} \, \|{\bf
			V}^{\perp}_{\B}\| \le  \mu_s \| {\bf V}^{\perp}_{\S} \| \\ \\
		\P^{\|}_{\B}(\pl ) - \, \ES \, {\bf V}^{\perp}_{\S}(\pl) + \, \left(1-\dfrac{\mu_s \|\velortS\|}{\|\velortB\|}\right)\, {\bf
			V}^{\perp}_{\B} & \,\, {\rm if} \,  \|{\bf
			V}^{\perp}_{\B}\| >  \mu_s \|{\bf V}^{\perp}_{\S}\|
	\end{array}
	\right.
\end{eqnarray}
This choice synthesizes the naive idea that, in the tangent direction, the
constraint $\B$ ``does its best'', in the sense that (\ref{impulso_attrito}) satisfies the following requirements:
\begin{itemize}
	\item if the components $\velortS({\bf p}_L )$ and $\velortB({\bf p}_L )$ give a velocity $\velort
	({\bf p}_L ) \= \velortS({\bf p}_L )+ \velortB({\bf p}_L )$ lying in the friction cone determined
	by $\mu_s$, then the unilateral constraint performs ``the whole'' tangent reaction, and
	$\velortB({\bf p}_R ) \= 0$. This means that the contact point of the system sticks to the unilateral constraint $\S$;

	\item if the tangent component $\velortB({\bf p}_L )$ is too big with respect to the orthogonal
	component $\velortS({\bf p}_L )$, so that $\velort ({\bf p}_L ) \= \velortS({\bf p}_L )+
	\velortB({\bf p}_L )$ does not lie in the friction cone determined by $\mu_s$, then the unilateral
	constraint performs ``the bigger possible'' tangent reaction, so that $\velortB({\bf p}_R ) \ne 0$ and the contact point slide on $\S$ in the direction of $\velortB({\bf p}_L )$.

\end{itemize}

A very similar characterization, even more affine to the rationale of the Morin--Amontons--Coulomb laws, introduces two different coefficients $\mu_s, \mu_d$ with $\mu_d \le \mu_s$ and a suitable definition of $\EB$ such that, in the end, the reactive impulse is given by
\begin{eqnarray}\label{impulso_attrito_din}
	{\bf I}(\pl) \= \left\{
	\begin{array}{lclcl}
		- \, (1+\ES) \, {\bf V}^{\perp}_{\S} \, -  \, {\bf
			V}^{\perp}_{\B} &\,& \,& {\rm if} \, \|{\bf
			V}^{\perp}_{\B}\| \le  \mu_s \| {\bf V}^{\perp}_{\S} \| \\ \\
		- \, (1+\ES) \, {\bf V}^{\perp}_{\S} \, - \, \dfrac{\mu_d \|\velortS\|}{\|\velortB\|} \, {\bf
			V}^{\perp}_{\B} &\, & \,& {\rm if} \,  \|{\bf
			V}^{\perp}_{\B}\| >  \mu_s \|{\bf V}^{\perp}_{\S}\|
	\end{array}
	\right.
\end{eqnarray}
The corresponding right velocity is then
\begin{eqnarray}\label{velocita_finale_attrito_din}
	\hskip-0.7truecm
	\pr \= \left\{
	\begin{array}{ll}
		\P^{\|}_{\B}(\pl ) - \, \ES \, {\bf V}^{\perp}_{\S}(\pl) & \,\, {\rm if} \, \|{\bf
			V}^{\perp}_{\B}\| \le  \mu_s \| {\bf V}^{\perp}_{\S} \| \\ \\
		\P^{\|}_{\B}(\pl ) - \, \ES \, {\bf V}^{\perp}_{\S}(\pl) + \, \left(1-\dfrac{\mu_d \|\velortS\|}{\|\velortB\|}\right)\, {\bf
			V}^{\perp}_{\B}(\pl) & \,\, {\rm if} \,  \|{\bf
			V}^{\perp}_{\B}\| >  \mu_s \|{\bf V}^{\perp}_{\S}\|
	\end{array}
	\right.
\end{eqnarray}

The proof of internal coherence requiring that, for null values of the friction coefficients, the rule gives back the non--ideal smooth constitutive characterization (\ref{caratt non_ideale 1}), is elementarily verified.
Moreover, the aforementioned characterizations (\ref{impulso_attrito}, \ref{impulso_attrito_din}) both satisfy the
requirement of energy consistency. Indeed, for the sake of brevity, let
$$
{\bf I}(\pl) \=	- \, (1+\ES) \, {\bf V}^{\perp}_{\S}(\pl) \, - \, \lambda \, {\bf
	V}^{\perp}_{\B}(\pl)
$$
with $\ES \in [0,1]$ and $\lambda \in [0,1]$ defined piecewise, and let
$$
\pl \=  \P^{\|}_{\B}(\pl)  + {\bf V}_{\S}^{\perp}(\pl) + {\bf V}_{\B}^{\perp}(\pl)) \, .
$$
For every frame of reference ${\mathbf h} \in \H_{\B}$, the following orthogonal
decompositions hold:
\begin{eqnarray*}
	\begin{array}{lcl}
		\pl - {\mathbf h} &\=&  \left(\P^{\|}_{\B}(\pl) - {\mathbf h}\right)  + {\bf V}_{\S}^{\perp}(\pl) + {\bf V}_{\B}^{\perp}(\pl)) \\
		\pr  - {\mathbf h} &\=&  \left(\P^{\|}_{\B}(\pl)  - {\mathbf h}\right)	- \ES \, {\bf V}^{\perp}_{\S}(\pl)  +  ( 1 - \lambda) \, {\bf V}^{\perp}_{\B}(\pl)
	\end{array}
\end{eqnarray*}
The balance of kinetic energy in the impact is therefore given by
\begin{eqnarray*}
	\Delta \KinH \= \KinH(\pr) - \KinH(\pl) \= - \, \dfrac{1 - \ES^2}{2} \, \| {\bf V}_{\S}^{\perp}(\pl)\|^2 \, - \,  \dfrac{1 - (1-\lambda)^2}{2} \, \| {\bf V}_{\B}^{\perp}(\pl)\|^2 \, ,
\end{eqnarray*}
which is non-positive for every impact, and vanishes if and only if $\ES \= 1, \, \lambda \= 0$.

\section{Examples}

\subsection{Example 1}

The simplest possible example is given by a material point $P$ of mass $m$ moving in a halfplane bordered by a rough horizontal line. 	The configuration space--time $\M$ can be locally described by introducing lagrangian coordinates $(t, x,y)$, where $x,y$ are the coordinates of $P$. The vertical metric $\Phi$ has then the local form $g_{ij} = diag(m,m)$. The contact condition of $P$ with the horizontal surface, that is the positional constraint $\S$, is given by $\S: y = 0$ (or alternatively $\S: y \ge 0$, in order to have a unilateral constraint).

The instantaneous kinetic constraint expressing the vanishing of the horizontal
velocity of the contact point $P$ is $\dot{x}= 0$. The RIGMS is then well defined (with $\A \= \JS$), and we have that
\begin{itemize}
	\item the elements of $\J$ have the local form
	$
	{\bf p} \= \dfrac{\partial}{\partial t} \, + \, \dot{x} \, \dfrac{\partial}{\partial x} \, + \, \dot{y} \, \dfrac{\partial}{\partial y} \, ;
	$
	\item the elements of $\JS\subset \J$ have the local form
	$
	{\bf p} \= \dfrac{\partial}{\partial t} \, + \, \dot{x} \, \dfrac{\partial}{\partial x}  \, ;
	$
	\item the  (unique) element of $\B \subset \JS$ is simply
	$
	{\bf p} \= \dfrac{\partial}{\partial t}  \, .
	$
\end{itemize}

Given a general left velocity
$
\pl \= \dfrac{\partial}{\partial t} \, + \, \dot{x}_L \, \dfrac{\partial}{\partial x} \, + \, \dot{y}_L \, \dfrac{\partial}{\partial y}
$
of the point, it is straightforward to show that
\begin{eqnarray*}
	\P^{\|}_{\B}(\pl ) \=  \dfrac{\partial}{\partial t} \, ; \qquad  {\bf V}^{\perp}_{\S}(\pl) \= \dot{y}_L \, \dfrac{\partial}{\partial y} \, ; \qquad {\bf V}^{\perp}_{\B}(\pl) \= \dot{x}_L \, \dfrac{\partial}{\partial x} \, .
\end{eqnarray*}
Once the coefficients $\ES, \mu_s$ (and possibly $\mu_d$) are known, using for example the constitutive characterization given by (\ref{impulso_attrito}), we obtain
\begin{eqnarray*}
	\hskip-0.4truecm
	\pr \= \left\{
	\begin{array}{ll}
		\dfrac{\partial}{\partial t} \, - \, \ES \,  \, \dot{y_L} \, \dfrac{\partial}{\partial y} & \,\, {\rm if} \, |\dot{x}_L| \le  \mu_s |\dot{y}_L | \\ \\
		\dfrac{\partial}{\partial t} \, + \, \left(1 - \dfrac{\mu_s |\dot{y}_L|}{|\dot{x}_L|}\right)  \, \dot{x}_L \, \dfrac{\partial}{\partial x} - \, \ES \,  \, \dot{y}_L \, \dfrac{\partial}{\partial y} & \,\, {\rm if} \, |\dot{x}_L| > \mu_s |\dot{y}_L |
	\end{array}
	\right.
\end{eqnarray*}
with immediately intuitive interpretation.

\subsection{Example 2}  A disk of radius $R$, mass $m$ and moment of inertia $A$ moves in a halfplane bordered by a rough horizontal line. 	The configuration space--time $\M$ can be locally described by lagrangian coordinates $(t, x,y, \vth)$, where $x,y$ are the
\begin{center} \label{FigDisk}
	\begin{minipage}[l]{.50\textwidth}
		\vskip0.2truecm
		\includegraphics[width=0.7\textwidth]{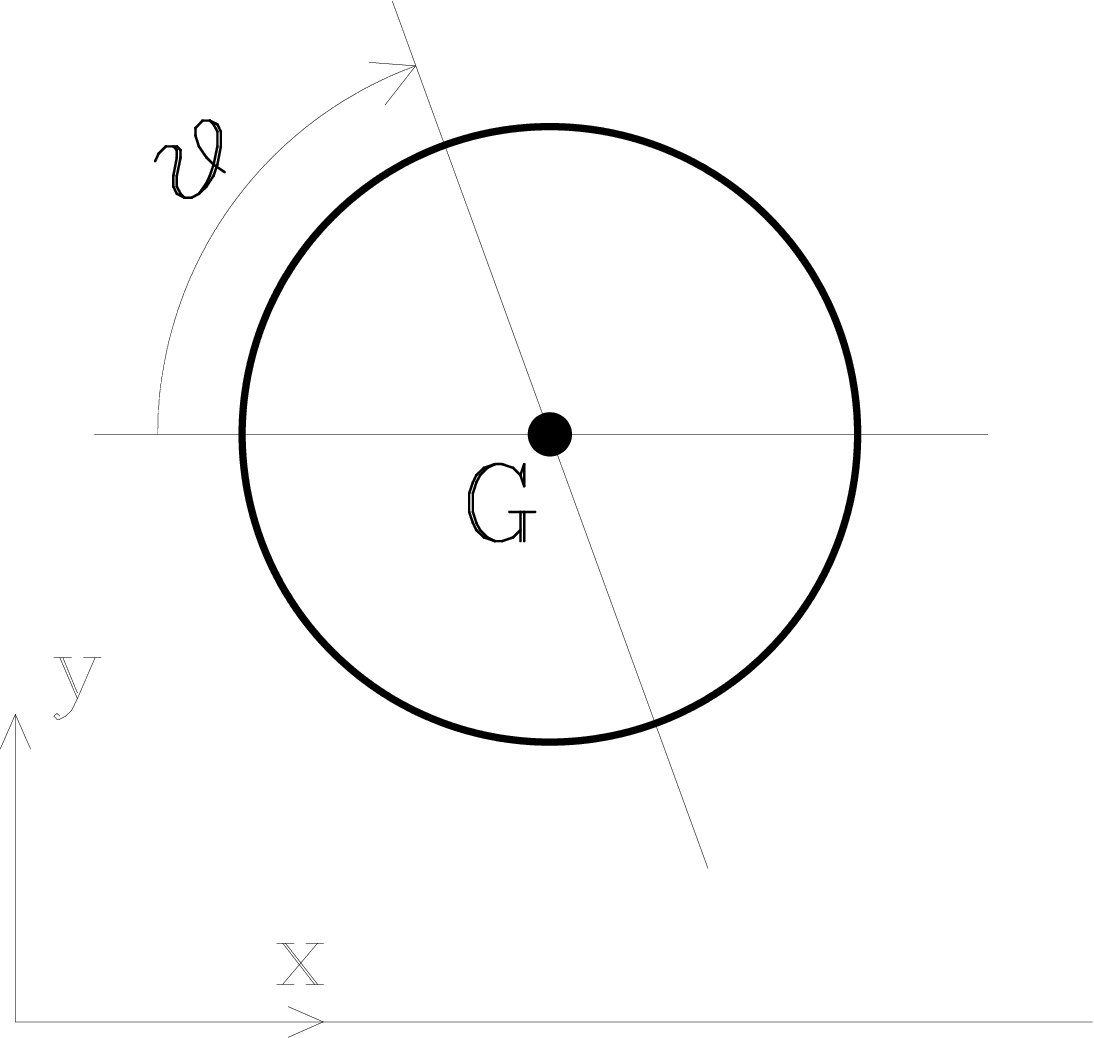}
		\vskip.1truecm		{\phantom{xxxxxxxxx} Fig.1}
	\end{minipage}%
	\hskip-0.8truecm	\begin{minipage}[r]{.56\textwidth}
		\vskip-0.0truecm   coordinates of the center of mass $G$ of the disk and $\vth$ the orientation of the disk (see Fig.1). The vertical metric $\Phi$ has then the local form $g_{ij} = diag(m,m,A)$.

		The contact condition of the disk with the horizontal surface, that is the positional constraint $\S$, is given by $\S: y - R = 0$ (or alternatively $\S: y - R \ge 0$).

		The instantaneous kinetic constraint expressing the vanishing of the horizontal velocity of the contact point $P$ is  $\dot{x} - R \dot{\vth} = 0$.
	\end{minipage}
\end{center}
The RIGMS is then well defined (with $\A \= \JS$), and we have that
\begin{itemize}
	\item the elements of $\J$ have the local form
	$$
	{\bf p} \= \dfrac{\partial}{\partial t} \, + \, \dot{x} \, \dfrac{\partial}{\partial x} \, + \, \dot{y} \, \dfrac{\partial}{\partial y}\, + \, \dot{\vth} \, \dfrac{\partial}{\partial \vth} \, ;
	$$
	\item the elements of $\JS\subset \J$ can be locally written as
	$$
	{\bf p} \= \dfrac{\partial}{\partial t} \, + \, \dot{x} \, \dfrac{\partial}{\partial x} \, + \, \dot{\vth} \, \dfrac{\partial}{\partial \vth} \, ;
	$$
	\item the elements of $\B \subset \JS$ can be locally written as
	$$
	{\bf p} \= \dfrac{\partial}{\partial t} \, + \, {R \dot{\vth}} \, \dfrac{\partial}{\partial x} \, + \,  \dot{\vth} \, \dfrac{\partial}{\partial \vth} \, .
	$$
\end{itemize}

Given a general left velocity
$
\pl \= \dfrac{\partial}{\partial t} \, + \, \dot{x}_L \, \dfrac{\partial}{\partial x} \, + \, \dot{y}_L \, \dfrac{\partial}{\partial y}\, + \, \dot{\vth}_L \, \dfrac{\partial}{\partial \vth}
$
of the system, straightforward but tedious calculations similar to those presented in \cite{MassPaga1997,Pasquero2004,Pasquero2005uni,Pasquero2006,Pasquero2008} give the following projections of $\pl$:
\begin{eqnarray*}
	\begin{array}{l}
		\begin{array}{lcl}
			\P^{\|}_{\B}(\pl ) =  \dfrac{\partial}{\partial t}  &+&  \dfrac{R}{mR^2 +A} \left(mR \dot{x}_L + A \dot{\vth}_L\right) \, \dfrac{\partial}{\partial x} \\ \\ &+& \dfrac{1}{mR^2 +A} \left(mR \dot{x}_L + A \dot{\vth}_L\right) \, \dfrac{\partial}{\partial \vth}
		\end{array}
		\\ \\
		\begin{array}{lcl}
			{\bf V}^{\perp}_{\S}(\pl) &\=& \dot{y}_L \, \dfrac{\partial}{\partial y}
		\end{array}
		\\ \\
		\begin{array}{lcl}
			{\bf V}^{\perp}_{\B}(\pl) &\=&  \dfrac{A}{mR^2 +A} \left(\dot{x}_L - R \dot{\vth}_L\right) \, \dfrac{\partial}{\partial x} \\ \\ && - \, \dfrac{mR}{mR^2 +A} \left(\dot{x}_L - R \dot{\vth}_L\right) \, \dfrac{\partial}{\partial \vth}
		\end{array}
	\end{array}
\end{eqnarray*}

Once the coefficients $\ES, \mu_s$ (and possibly $\mu_d$) are known, then (\ref{impulso_attrito}) gives the following two possibilities.

\noindent If $|\dot{x}_L - R \dot{\vth}_L| \, \sqrt{\dfrac{A}{mR^2+A}}  \le  \mu_s |\dot{y}_L |$, then
\begin{eqnarray*}
	\begin{array}{lcl}
		\pr \= \dfrac{\partial}{\partial t}  &+&  \dfrac{R}{mR^2 +A} \left(mR \dot{x}_L + A \dot{\vth}_L\right) \, \dfrac{\partial}{\partial x}  -  \ES \,  \, \dot{y_L} \, \dfrac{\partial}{\partial y} \\ \\
		&& + \dfrac{1}{mR^2 +A} \left(mR \dot{x}_L + A \dot{\vth}_L\right) \, \dfrac{\partial}{\partial \vth}
	\end{array}
\end{eqnarray*}

\noindent Otherwise, if $ |\dot{x}_L - R \dot{\vth}_L| \, \sqrt{\dfrac{A}{mR^2+A}}  >  \mu_s |\dot{y}_L |$, then
\begin{eqnarray*}
	\begin{array}{lcl}
		\pr \= \dfrac{\partial}{\partial t}  &+&  \left[\dot{x}_L - \dfrac{\mu_s |\dot{y}_L |}{ |\dot{x}_L - R \dot{\vth}_L| \, \sqrt{\dfrac{A}{mR^2+A}}} \, \dfrac{A}{mR^2+A}\left( \dot{x}_L - R\dot{\vth}_L\right) \right] \, \dfrac{\partial}{\partial x} \\ \\ && -  \ES \,  \, \dot{y_L} \, \dfrac{\partial}{\partial y} \\ \\
		&& +  \left[\dot{\vth}_L + \dfrac{\mu_s |\dot{y}_L |}{|\dot{x}_L - R \dot{\vth}_L| \, \sqrt{\dfrac{A}{mR^2+A}}} \, \dfrac{mR}{mR^2+A}\left( \dot{x}_L - R\dot{\vth}_L\right) \right] \, \dfrac{\partial}{\partial \vth} \,.
	\end{array}
\end{eqnarray*}
The results about the impact of the disk is clearly much less easily interpretable with respect to that of the material point. However, three particular initial data show its reliability:
\begin{itemize}
	\item if the initial velocity $\pl$ of the disk at the impact is such that $\dot{x}_L - R \dot{\vth}_L \= 0$, then $	{\bf V}^{\perp}_{\B}(\pl)\= 0$, the frictional kinetic constraint $\B$ does not act on the horizontal and angular velocities of the disk, and the only  action of the constraint in the impact is to reflect (partially) the orthogonal velocity $\velortS$;

	\item if the initial velocity $\pl$ of the disk at the impact is such that $\dot{x}_L \= 0, \dot{\vth}_L >0$, then the frictional kinetic constraint $\B$ acts on both the linear and angular velocities of the disk. The reaction of the constraint in the impact reflects (partially) the orthogonal velocity $\velortS$, gives a positive horizontal component to the exit velocity and dampens the rotation of the disk (with a symmetrical behavior if $\dot{x}_L \= 0, \dot{\vth}_L < 0$);

	\item if the initial velocity $\pl$ of the disk at the impact is such that $\dot{x}_L > 0, \dot{\vth}_L \= 0$, then the frictional kinetic constraint $\B$ acts on both the linear and angular velocities of the disk. The reaction of the constraint in the impact reflects (partially) the orthogonal velocity $\velortS$, cuts down the horizontal component of the exit velocity and gives a positive component to the rotation of the disk (with a symmetrical behavior if $\dot{x}_L < 0, \dot{\vth}_L = 0$).
\end{itemize}

\subsection{Example 3} A rod of lenght $2L$, mass $m$ and moment of inertia $A$ moves in a halfplane bordered by a rough horizontal line. 	The configuration space--time $\M$ can be locally described by lagrangian coordinates $(t, x,y, \vth)$, where  $x,y$ are the
\begin{center} \label{FigRod}
	\begin{minipage}[l]{.50\textwidth}
		\vskip0.3truecm
		\includegraphics[width=0.8\textwidth]{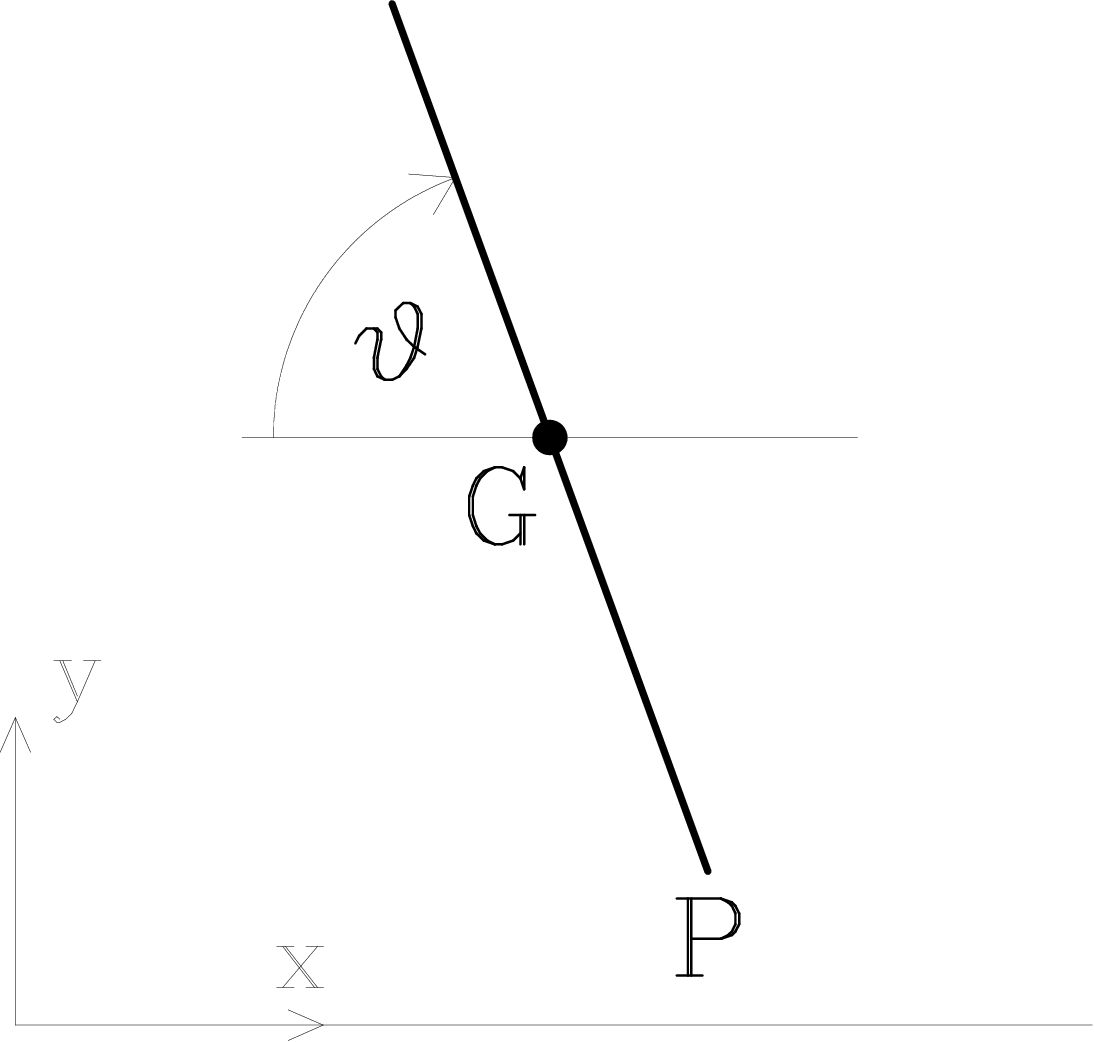}
		\vskip.1truecm		{\phantom{xxxxxxxxx} Fig.2}
	\end{minipage}%
	\hskip-0.8truecm	\begin{minipage}[r]{.56\textwidth}
		\vskip-0.0truecm   coordinates of the center of mass $G$ of the rod and $\vth$ the orientation of the rod (see Fig.2). The vertical metric $\Phi$ has then the local form $g_{ij} = diag(m,m,A)$. For $\vth \in (0,\pi)$, the contact condition of the extreme $P$ with the horizontal surface, that is the positional constraint $\S$, is given by $\S: y - L \sin\vth = 0$ (or alternatively $\S: y - L \sin\vth \ge 0$).

		The instantaneous kinetic constraint expressing the vanishing of the horizontal velocity of the contact point $P$ is $\dot{x} - L \dot{\vth} \sin\theta = 0$.
	\end{minipage}
\end{center}
The RIGMS is then well defined (with $\A \= \JS$), and we have that
\begin{itemize}
	\item the elements of $\J$ have the local form
	$$
	{\bf p} \= \dfrac{\partial}{\partial t} \, + \, \dot{x} \, \dfrac{\partial}{\partial x} \, + \, \dot{y} \, \dfrac{\partial}{\partial y}\, + \, \dot{\vth} \, \dfrac{\partial}{\partial \vth} \, ;
	$$
	\item the elements of $\JS\subset \J$ can be locally written as
	$$
	{\bf p} \= \dfrac{\partial}{\partial t} \, + \, \dot{x} \, \dfrac{\partial}{\partial x} \, + \, {L \dot{\vth} \cos \vth} \, \dfrac{\partial}{\partial y}\, + \, \dot{\vth} \, \dfrac{\partial}{\partial \vth} \, ;
	$$
	\item the elements of $\B \subset \JS$ can be locally written as
	$$
	{\bf p} \= \dfrac{\partial}{\partial t} \, + \, {L \dot{\vth} \sin \vth} \, \dfrac{\partial}{\partial x} \, + \, {L \dot{\vth} \cos \vth} \, \dfrac{\partial}{\partial y}\, + \, \dot{\vth} \, \dfrac{\partial}{\partial \vth} \, .
	$$
\end{itemize}

Given a general left velocity
$
\pl \= \dfrac{\partial}{\partial t} \, + \, \dot{x}_L \, \dfrac{\partial}{\partial x} \, + \, \dot{y}_L \, \dfrac{\partial}{\partial y}\, + \, \dot{\vth}_L \, \dfrac{\partial}{\partial \vth} \, ,
$
calculations similar to those presented in \cite{MassPaga1997,Pasquero2004,Pasquero2005uni,Pasquero2006,Pasquero2008} give the following projections of $\pl$:
\begin{eqnarray*}
	\begin{array}{l}
		\begin{array}{lcl}
			\P^{\|}_{\B}(\pl ) =  \dfrac{\partial}{\partial t}  &+&  \dfrac{L \sin \vth}{mL^2 +A} \left(mL\dot{x}_L \sin \vth + mL\dot{y}_L \cos \vth + A \dot{\vth}_L\right) \, \dfrac{\partial}{\partial x} \\ \\  &+& \dfrac{L \cos \vth}{mL^2 +A} \left(mL\dot{x}_L \sin \vth + mL\dot{y}_L \cos \vth + A \dot{\vth}_L\right) \, \dfrac{\partial}{\partial y} \\ \\ &+& \dfrac{1}{mL^2 +A} \left(mL\dot{x}_L \sin \vth + mL\dot{y}_L \cos \vth + A \dot{\vth}_L\right) \, \dfrac{\partial}{\partial \vth}
		\end{array}
	\end{array}
\end{eqnarray*}

\begin{eqnarray*}
	\begin{array}{l}
		\begin{array}{lcl}
			{\bf V}^{\perp}_{\S}(\pl) &\=& \dfrac{A}{mL^2 \cos^2 \vth +A} \,\left(\dot{y}_L - L \dot{\vth}_L \cos \vth\right) \, \dfrac{\partial}{\partial y} \\ \\
			&&  - \, \dfrac{mL \cos \vth}{mL^2 \cos^2 \vth +A} \,\left(\dot{y}_L - L \dot{\vth}_L \cos \vth\right) \,  \dfrac{\partial}{\partial \vth}
		\end{array}
		\\ \\
		\begin{array}{lcl}
			{\bf V}^{\perp}_{\B}(\pl) &=&  \dfrac{mL^2 \cos^2\vth +A}{mL^2 +A} \left[\dot{x}_L - \dfrac{L \sin \vth}{mL^2 \cos^2 \vth +A} \left(mL\dot{y}_L \cos \vth + A \dot{\vth}_L\right)\right] \, \dfrac{\partial}{\partial x} \\ \\
			&& - \, \dfrac{mL^2 \sin\vth \cos\vth}{mL^2 +A} \left[\dot{x}_L - \dfrac{L \sin \vth}{mL^2 \cos^2 \vth +A} \left(mL\dot{y}_L \cos \vth + A \dot{\vth}_L\right)\right] \, \dfrac{\partial}{\partial y} \\ \\
			&& - \, \dfrac{mL \sin\vth}{mL^2 +A} \left[\dot{x}_L - \dfrac{L \sin \vth}{mL^2 \cos^2 \vth +A} \left(mL\dot{y}_L \cos \vth + A \dot{\vth}_L\right)\right] \,  \dfrac{\partial}{\partial \vth}
		\end{array}
	\end{array}
\end{eqnarray*}

The reader can determine the general expression of the right velocities $\pr$ in the cases ${\| {\bf V}^{\perp}_{\B}(\pl) \|} \le \mu_s {\| {\bf V}^{\perp}_{\S}(\pl) \|}$ and ${\| {\bf V}^{\perp}_{\B}(\pl) \|} > \mu_s {\| {\bf V}^{\perp}_{\S}(\pl) \|}$. Once again, a qualitative analysis of the behavior of the rod for some specific impacts gives significant results. For instance, we have that:
\begin{itemize}
	\item if the rod impacts the line vertically, with $\dot{x}_L=0, \dot{y}_L<0, \dot{\vth}_L =0$ and $\vth \= \pi/2$, so that $\pl \= \dfrac{\partial}{\partial t} + \dot{y}_L \, \dfrac{\partial}{\partial y}$ and $\vth \= \pi/2$, we have ${\bf V}^{\perp}_{\B}(\pl)\= 0 $ and then $\pr \= \dfrac{\partial}{\partial t} - \ES \, \dot{y}_L \, \dfrac{\partial}{\partial y}$ independently of the friction coefficient $\mu_s$. So the rebound preserves the verticality;
	\item if the rod falls vertically and inclined, so that $\dot{x}_L=0, \dot{y}_L<0, \dot{\vth}_L =0$ with $\vth \ne \pi/2$, and ${\| {\bf V}^{\perp}_{\B}(\pl) \|} \le \mu_s {\| {\bf V}^{\perp}_{\S}(\pl) \|}$, we have that:
	\begin{eqnarray*}
		\begin{array}{lcl}
			\pr &\=& \dfrac{\partial}{\partial t} + \dfrac{mL^2 \sin\vth \cos\vth}{mL^2 +A} \, \dot{y}_L \, \dfrac{\partial}{\partial x} \\ \\
			&& \quad + \left(\dfrac{mL^2\cos^2\vth}{mL^2 +A} - \ES \dfrac{A}{mL^2\cos^2\vth + A}\right) \dot{y}_L \, \dfrac{\partial}{\partial y}\\ \\
			&& \quad \quad + \left(\dfrac{mL\cos\vth}{mL^2 +A} + \ES \dfrac{mL\cos\vth}{mL^2\cos^2\vth + A}\right) \dot{y}_L \, \dfrac{\partial}{\partial \vth}
		\end{array}
	\end{eqnarray*}
	In particular, we have that the center of the rod assumes
	\begin{itemize}
		\item a horizontal component $\dot{x}_R$ of the right velocity having the opposite sign of $\cos\vth$;
		\item a vertical component $\dot{y}_R$ of the right velocity having a sign depending on the impact angle $\vth$ and the restitution coefficient $\ES$. In particular, the rod rebounds with a positive vertical velocity if and only if $$\cos^2\vth < \dfrac{A}{2mL^2}\left(\sqrt{1+\dfrac{4 \ES (mL^2 +A)}{A}}-1\right) \, ;$$
		\item an angular velocity $\dot{\vth}_R$ of the right velocity having the opposite sign of $\cos\vth$.
	\end{itemize}
	Note however that the condition ${\| {\bf V}^{\perp}_{\B}(\pl) \|} \le \mu_s {\| {\bf V}^{\perp}_{\S}(\pl) \|}$ depends only on the inertial properties $m,A$ of the rod, on the friction coefficient $\mu_s$ and on the impact angle $\vth$, but not on the impact velocity $\dot{y}_L$;
	\item if the rod falls vertically and inclined, so that $\dot{x}_L=0, \dot{y}_L<0, \dot{\vth}_L =0$ with $\vth \ne \pi/2$, and ${\| {\bf V}^{\perp}_{\B}(\pl) \|} > \mu_s {\| {\bf V}^{\perp}_{\S}(\pl) \|}$, choosing the characterization (\ref{impulso_attrito}), we have that:
	\begin{eqnarray*}
		\hskip-1truecm
		\begin{array}{ll}
			\pr & \= \dfrac{\partial}{\partial t} + \dfrac{ \mu_s {\| {\bf V}^{\perp}_{\S}(\pl) \|}}{\| {\bf V}^{\perp}_{\B}(\pl) \|} \, \dfrac{mL^2 \sin\vth \cos\vth}{mL^2 +A} \, \dot{y}_L \, \dfrac{\partial}{\partial x} \\ \\
			\, + & \left(\dfrac{mL^2\cos^2\vth - \ES A}{mL^2\cos^2\vth +A} - \, \dfrac{ \mu_s {\| {\bf V}^{\perp}_{\S}(\pl) \|}}{\| {\bf V}^{\perp}_{\B}(\pl) \|} \, \dfrac{m^2L^4 \sin^2\vth \cos^2\vth}{(mL^2 +A)(mL^2\cos^2\vth +A)} \right) \dot{y}_L \, \dfrac{\partial}{\partial y}\\ \\
			\, + & \left(\dfrac{mL\cos\vth}{mL^2\cos^2\vth + A} \left(1+ \ES - \dfrac{ \mu_s {\| {\bf V}^{\perp}_{\S}(\pl) \|}}{\| {\bf V}^{\perp}_{\B}(\pl) \|} \, \dfrac{mL^2 \sin^2\vth}{mL^2 +A} \right)  \right) \dot{y}_L \, \dfrac{\partial}{\partial \vth}
		\end{array}
	\end{eqnarray*}
	In particular, we have that the center of the rod assumes
	\begin{itemize}
		\item a horizontal component $\dot{x}_R$ of the right velocity having the same sign of $\cos\vth$ (the rod ``slips forward'' in the direction where it is inclined);
		\item a vertical component $\dot{y}_R$ of the right velocity that can change sign depending on the angle $\vth$  of impact and on the values of restitution and friction coefficients $\ES, \mu_s$;
		\item an angular velocity $\dot{\vth}_R$ of the right velocity having the opposite sign of $\cos\vth$.
	\end{itemize}
\end{itemize}

\section{Conclusions and potential developments}

We present a model for the study of frictional impacts of mechanical systems with unilateral rough constraints based only on the structure and properties of the geometric framework of the space--time bundle associated to a mechanical system.

The model intentionally does not involve analyses of contact forces and relies instead on the knowledge of the geometry of the constraint, the kinetic data of the system at the instant of the impact, and the assignment of two (possibly three) coefficients dependent on the nature of the contact. The frictional action of the unilateral constraint is modeled by the impulsive action of a breakable instantaneous kinetic  constraint expressing the sticking of the contact point of the system with the constraint. The rule determining the behavior of the system after the impact finds its rationale in the Amontons-Coulomb-Morin laws for dry friction carried in the context of the impulsive systems.

The method allows natural applications to similar but not identical problems, such as the impact between two rigid bodies, simply by expressing the contact of the bodies in form of unilateral constraint. The simplest example is given by the impact of two spheres modeled with the constraint expressing the distance of their centers.

The model gives also the first step to approach in a geometric way the problem of multibody impacts with friction. A possible way forward is analogous to the one that lead from the geometry of single point impacts without friction (see \cite{Pasquero2005uni}) to that of multi--point impacts without friction (see \cite{pasquero2018MIl}).

Moreover, the same model could be applied to analyse from a different point of view all the problems of constrained mechanical systems with assigned initial data in presence of friction, such as the Classical Painlev\'e Paradox, viewed as impacts with a kinetic constraint, possibly in the context of inert constraints.

	\bibliographystyle{unsrt}
	\bibliography{bibart,biblib}

\end{document}